\documentclass[runningheads]{llncs}
\usepackage{graphicx}
\usepackage{multirow}
\usepackage{comment}
\usepackage{float}
\usepackage{subfig}
\usepackage{color}
\usepackage{amsmath}
\usepackage{booktabs}
\usepackage{amsfonts}

\begin{document}
\title{ICA-UNet: ICA Inspired Statistical UNet for Real-time 3D Cardiac Cine MRI Segmentation}
\titlerunning{ICA-UNet for Real-time 3D Cardiac Cine MRI Segmentation}
%
\author{
Tianchen Wang\thanks{Both contributed equally.}\inst{1} \and 
Xiaowei Xu$^\star$\inst{2} \and 
Jinjun Xiong\inst{3} \and
Qianjun Jia\inst{2} \and
Haiyun Yuan\inst{2} \and 
Meiping Huang\inst{2} \and 
Jian Zhuang\inst{2} \and
Yiyu Shi\inst{1} 
}
\authorrunning{T.Wang, et al.}
\institute{University of Notre Dame, \\
\email{\{twang9, yshi4\}@nd.edu}
\and Guangdong Provincial People's Hospital\\
\email{xiao.wei.xu@foxmail.com, jiaqianjun@126.com, yhy\_yun@163.com, huangmeiping@126.com, zhuangjian5413@tom.com}
\and IBM Thomas J. Watson Research Center, \\ 
\email{jinjun@us.ibm.com}
}
\maketitle               
\begin{abstract}
Real-time cine magnetic resonance imaging (MRI) plays an increasingly important role in various cardiac interventions. In order to enable fast and accurate visual assistance, the temporal frames need to be segmented on-the-fly. However, state-of-the-art MRI segmentation methods are used either offline because of their high computation complexity, or in real-time but with significant accuracy loss and latency increase (causing visually noticeable lag). As such, they can hardly be adopted to assist visual guidance. In this work, inspired by a new interpretation of Independent Component Analysis (ICA) \cite{hyvarinen2000independent} for learning, we propose a novel ICA-UNet for real-time 3D cardiac cine MRI segmentation. Experiments using the MICCAI ACDC 2017 dataset show that, compared with the state-of-the-arts, ICA-UNet not only achieves higher Dice scores, but also meets the real-time requirements for both throughput and latency (up to 12.6$\times$ reduction), enabling real-time guidance for cardiac interventions without visual lag. 
\end{abstract}

\section{Introduction}
Real-time cine Magnetic Resonance Imaging (MRI) has enabled fast and accurate visual guidance in various cardiac interventions, such as
aortic valve replacement \cite{mcveigh2006real}, 
cardiac electroanatomic mapping and ablation \cite{radau2011vurtigo}, 
electrophysiology for atrial arrhythmias \cite{vergara2011real}, 
intracardiac catheter navigation \cite{gaspar2014three}, and
myocardial chemoablation \cite{rogers2016transcatheter}.
In these applications, it is strongly desirable to segment
the temporal frames on-the-fly, satisfying both throughput and latency requirements. 
The throughput should be at least above the cine MRI reconstruction rate of 
22 frames per second (FPS) \cite{schaetz2017accelerated,iltis2015high}. The latency should be 
no more than 50 ms to avoid visually noticeable lags \cite{annett2014}. 
Most of the existing segmentation methods \cite{isensee2017automatic,zotti2018convolutional,yan2018left,vigneault2018omega,xu2019whole,xu2018quantization}, however, focus on accuracy. In order to handle 
cardiac border ambiguity and large variations among target objects from different patients, these methods come with high computation cost.
Hence their inference latency and throughput are far from meeting the real-time requirements and thus can only be applied offline.


MSU-Net \cite{wang2019msu} was proposed in MICCAI'19 as the first framework achieving the real-time segmentation of 3D cardiac cine MRI. It uses a canonical form distribution to describe the multiple input frames in a snippet of cine MRI so that only a single pass through the network is needed for all the frames in the snippet. While MSU-Net increases the throughput drastically, the inference latency 
is also increased to well above 50 ms due to the need of input clustering, i.e., the inference is carried out only after all the frames in a snippet have arrived. When MSU-Net is applied to real-time cine MRI segmentation, such significant visual lags jeopardize the effectiveness of visual guidance in cardiac intervention.  

As a popular computational method for
decomposing a multivariate signal into additive  
independent non-Gaussian signals (bases),
Independent Component Analysis (ICA) has been widely used in multiple image processing applications such as noise reduction \cite{hyvarinen2000independent},
image separation in medical data \cite{delorme2007enhanced,bronstein2005sparse} and image decomposition \cite{starck2005image}. 
Through the unmixing process in ICA, 
any image patch out of a given image
can be represented by a linear combination 
of a set of independent bases of the same size as the image patch. 
In the mixing process, the original image can be reconstructed using the 
bases with proper coefficients. 

In this paper, based on a new interpretation of ICA for learning (Section~\ref{sec:motiv}), 
we propose ICA-UNet, a novel model that can not only achieve highly accurate 
3D cardiac cine MRI segmentation results,
but also attain both high throughput and low latency.
Specifically, an input temporal frame in the cine MRI is decomposed into independent
bases and a mixing tensor, composed of the coefficient tensors of all the bases, by a light-weight 
ICA-encoder. Such an ICA-encoder mimics the unmixing operation in ICA. A U-Net like architecture 
is trained to learn the transform of the mixing tensor 
from its original function of image reconstruction to the target function of 
image segmentation. As such, the transformed
mixing tensors can be mixed with the bases through light-weight ICA-decoders to get the desirable features for final segmentation evaluation. 
Because the coefficient tensors that compose the mixing tensor are much smaller in size than the input frame and can be processed in parallel
due to the independence between their corresponding bases, significant latency reduction can be achieved. 

Experiment results show that, compared with the state-of-the-art real-time cardiac cine MRI segmentation method MSU-Net, ICA-UNet achieves much higher Dice scores for all cardiac classes with up to 12.6$\times$ latency reduction. More specifically, the latency of ICA-UNet is below 50 ms while its throughput is still above 22 FPS, which implies that ICA-UNet is the first method 
meeting the real-time performance requirements in terms of both throughput and latency for
MRI guided cardiac intervention with no visually noticeable lags. 
In fact, the accuracy achieved by ICA-UNet is on a par with state-of-the-art methods that focus on accuracy and can only run offline because of their complexity.


\section{Motivation}
\label{sec:motiv}
It has long been recognized that Independent Component Analysis (ICA) can be used to
extract features (bases) from images \cite{hoyer2000independent}. 
Following the similar
setup as \cite{olshausen1996natural}, we can
partition an image into a set of smaller image patches such that
each image patch can be represented
as a linear combination of independent basis image
patches along with their coefficients. Compactly put, 
for a set of input image patches $D$ where
each row vector of $D$ represents an input
image patch,
the goal of ICA is to estimate the unmixing matrix $W$ such that
the realizations of bases $X=WD$
are as mutually independent as possible (which is called the unmixing process), while
the reconstruction of input
image patches, $AX$, is
as close to $D$ as possible (which is called the mixing process). 
Matrix $A$ is called the
mixing matrix, which equals the pseudo
inverse of $W$. 
There are different ICA algorithms and implementations, and one popular  
implementation is FastICA\cite{hyvarinen2000independent}. 
In the rest of the paper, we extend the matrix term to tensor as mixing tensor $\mathsf{A}$ 
and basis tensor $\mathsf{X}$, 
due to the multi-dimension nature of the input images.  Each channel in $\mathsf{X}$ represents 
a basis and the number of channels $m$ equals the basis dimension of ICA. The corresponding coefficient 
tensor of each basis can be extracted from the mixing tensor $\mathsf{A}$.
We will also use basis tensor and bases interchangeably for the simplicity of discussion.   

\begin{figure}[!t]
\begin{center}
\includegraphics[width=1.0\linewidth]{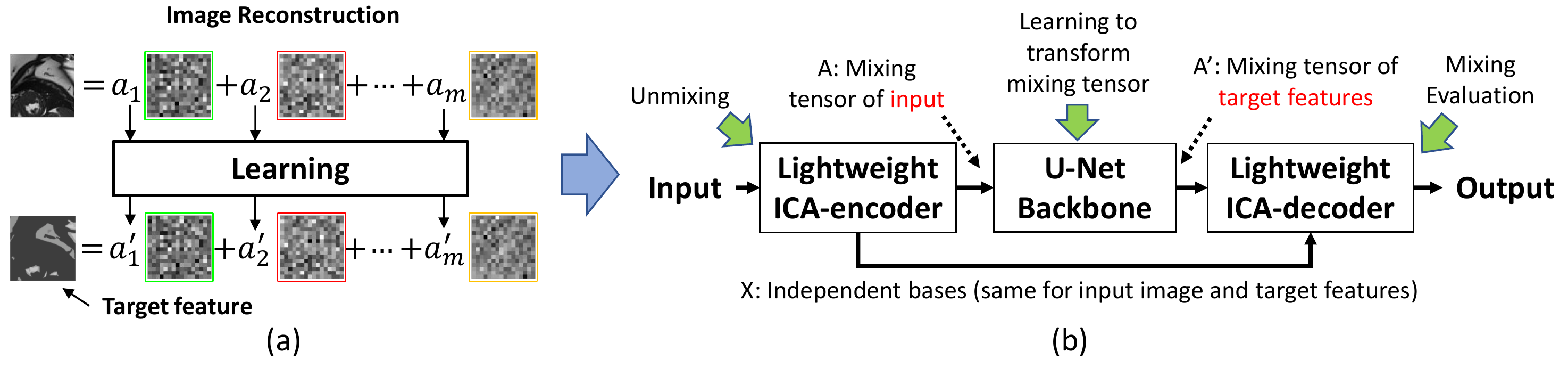}
\end{center}
\caption{(a) Conceptual illustration of the mixing tensor transform. 
We desire to learn the transform of the mixing tensor (formed by the coefficients $a_1$, 
$a_2$,...$a_m$), from the initial reconstruction task where it is extracted, to the target task. During the transform these coefficients can be processed in parallel due to basis independence.  
The bases will be saved and reused with the transformed mixing tensor to obtain target features.
(b) Conceptual illustration of ICA-UNet. An encoder-backbone-decoder design where the lightweight ICA-encoder/decoder mimic the ICA unmixing/mixing operations and the U-Net learns the transform of the mixing tensor. The bases sharing between ICA-encoder/decoder also naturally forms a long-range skip connection which helps information traverse. }
\label{fig:icalearn}
\end{figure}

We can have an interesting interpretation of ICA: 
Both the mixing tensor and the bases can be considered as some kind of feature representation of the input image, with
the bases being more fundamental
to the input image while the mixing tensor
being more related to a particular application such
as reconstruction of the input image. In other
words, bases can be treated
as well-behaved image features that can be reused
for different applications, while the mixing tensor
can be treated as weights of a simple fully connected
layer used to reconstruct the input image.

With such an insight, we wonder if we can learn to transform the mixing tensor so that it can be utilized for a different set of applications (with the help of the bases) beyond the original image reconstruction.
As illustrated in Fig.~\ref{fig:icalearn}(a), 
the original mixing tensor for 
image reconstruction is transformed to that for target application, which, 
after mixing together with the bases, can be used to get the desirable target features for final evaluation of the target application. 
As the bases are shared, only the mixing tensor, which is composed of the coefficient tensors of all the bases, needs to be transformed. During this process, 
the coefficient tensors of different bases can be computed in parallel 
due to the independence between the bases, and each of which is much smaller than the original input. Thus significant latency reduction can be achieved. 
Since the mixing tensor still exhibits spatial patterns, 
a conventional image oriented deep 
neural networks such as U-Net can be used as the backbone to learn the transform.

In conventional ICA, the unmixing operation is lossy,
which affects the downstream application (such as image reconstruction) accuracy,
and runs as a separate optimization process,
which can be 
quite time-consuming. Since
the learning of the target mixing tensor is
also an optimization problem, why not 
combining the ICA process
with the learning process as a joint
end-to-end training process so that
we can not only mitigate the impact
of lossy unmixing operation on accuracy, but also reduce one separate optimization process?
Such a motivation drives us to
propose a lightweight neural network based ICA encoder and decoder to 
mimic the unmixing and mixing operations in ICA. We further integrate 
them with a U-Net backbone so that they can be end-to-end trained.









\section{Method}
\begin{figure}[!t]
\begin{center}
\includegraphics[width=1\linewidth]{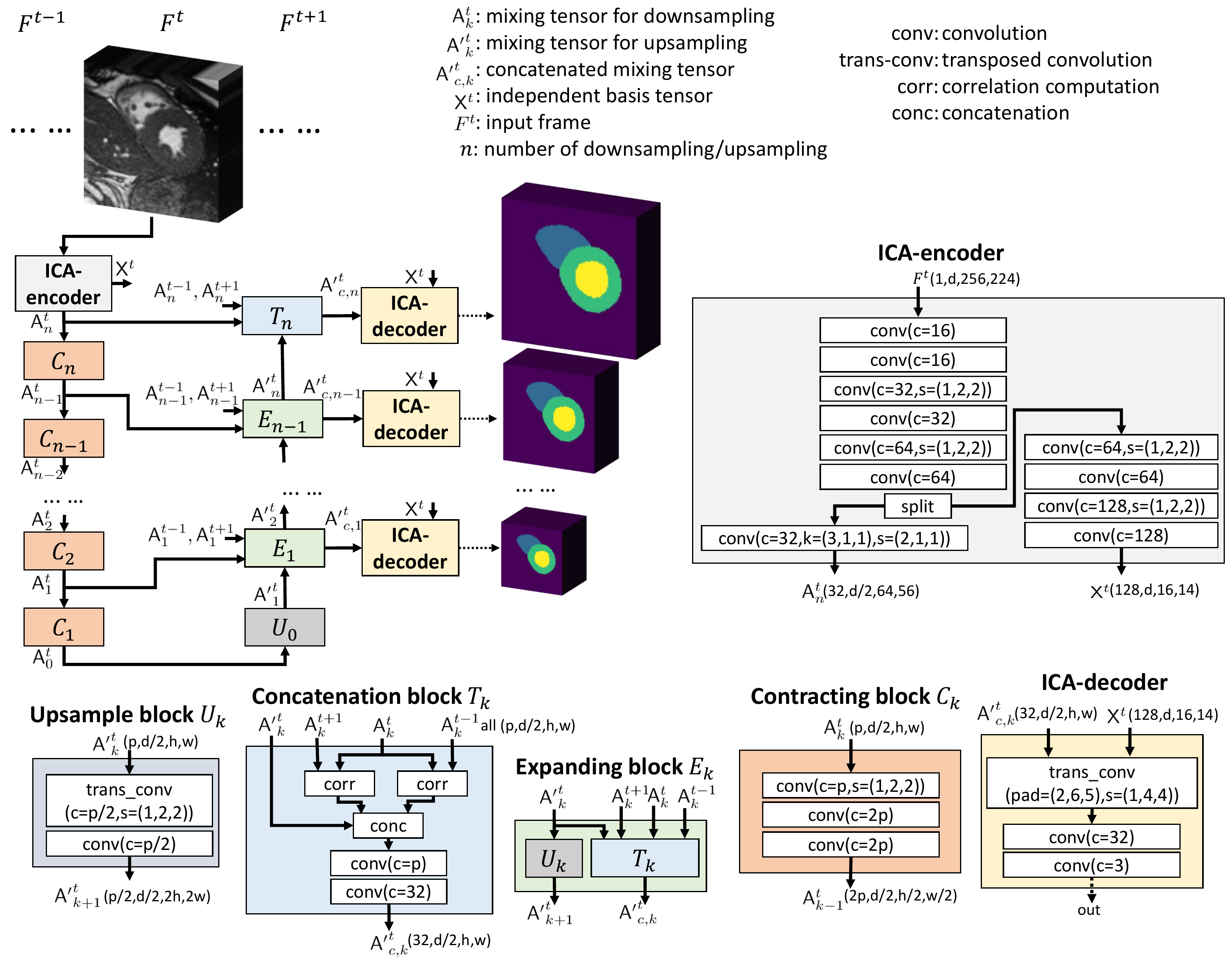}
\end{center}
\caption{
The architecture of ICA-UNet. 
The detailed structure of ICA-UNet. p in dimension representation denotes the channel width of input features. Unless specified, the default configuration for conv block (convolution batch normalization and leaky ReLU appended) is kernel\_size(k)=(1,3,3), stride(s)=(1,1,1), padding=(0,1,1) with output channel width c, and the default configuration for trans\_conv block (transposed convolution) is kernel\_size=(1,2,2), stride(s)=(1,2,2), padding(pad)=(0,0,0) with output channel width c.
}
\label{fig:ica_unet}
\end{figure}

Driven by the motivation in Section~\ref{sec:motiv}, a conceptual illustration of our ICA-UNet is shown in Fig.~\ref{fig:icalearn}(b). Its detailed architecture is shown in Fig.~\ref{fig:ica_unet}, where we
use superscript to denote the time stamp of input frame (e.g. $F^t$). 
A summary of all the notations used 
in this section is also included in the figure. 
ICA-UNet is mainly made of four types of modules:
the ICA-encoder, 
the contracting blocks ($C_k, k\!\in\![1, n]$), the expanding blocks ($E_k, k\!\in\![1, n]$),
and the ICA-decoders. $n$ is the number of contracting/expanding blocks acting as a hyperparameter that can affect accuracy and speed, as will be shown in the experiments.

\noindent\textbf{1) ICA-encoder: }
The ICA-encoder 
extracts both the statistically independent basis tensor
and the associated initial mixing tensor from the input image. 
Instead of running a standard ICA
process where the image needs to be explicitly partitioned
and an explicit iterative optimization is needed to
obtain the mixing tensor
and the realization of the bases, 
we propose to use a neural network to obtain them as a function of the input image.

For an input frame size of ($d$,$h$,$w$) for depth, height, and width, respectively, we can choose the basis dimension $m$ 
(in this paper $m=32$), 
the size of the initial mixing tensor $\mathsf{A}_n^t$ as (1,$m$,$d/2$,$h/4$,$w/4$), 
and the size of independent basis tensor $\mathsf{X}^t$ as (1,$u\!\times \!m$,$d$,$h/16$,$w/16$),
where $u$ is the output channel width of the transposed convolution between the concatenated mixing tensor $\mathsf{A'}_{c,k}^t$ and $\mathsf{X}^t$ in the ICA-decoder ($u\!=\!4$ in this paper). 
Each channel in $\mathsf{X}^t$ corresponds to an independent basis. The corresponding coefficient tensor of each basis can be 
obtained from the mixing tensor $\mathsf{A}_n^t$, each of size (1,1,$d/2$,$h/4$,$w/4$). The extraction of $\mathsf{X}^t$ and $\mathsf{A}_n^t$
shares some layers for low level feature extraction before they are 
split channel-wise in order 
to reduce the computation. 

After both $\mathsf{A}_n^t$ and $\mathsf{X}^t$ are obtained, $\mathsf{A}_n^t$ is forwarded to the following
contracting block $C_n$, 
and $\mathsf{X}^t$ is directly forwarded to each ICA-decoder block for mixing operation.

The objective function of ICA-encoder, which is used for regulating the optimization towards sparsity, independence and accuracy, can be expressed as 
\begin{equation}
{\textbf{min}} ~\mathcal{L}_{\text{ICA}}=\lambda_s||(\mathsf{X}^t)||_1+\lambda_i\text{avg}(-\alpha\log\cosh(\mathsf{X}^t/\alpha))+\lambda_r\!\parallel\!{\mathsf{A}_n^t\otimes\mathsf{X}^t\!-\!F^t}\!\parallel_2^2,
\label{eq:icanet_loss}
\end{equation}
where $\lambda_s$, $\lambda_i$ and $\lambda_r$ are the weights of the loss terms which are set to 1.0 in our experiments. 
The first term reflects sparsity through L1 norm. 
The second term reflects independence through neg-entropy \cite{hyvarinen2004independent}; $\alpha$ is a constant number between $0.5$ and $1$ (we take $\alpha=0.75$ in our experiments); $\text{avg}(\cdot)$ denotes 
element-wise average. 
The third term reflects reconstruction loss. We adopt transposed convolution ($\otimes$) as the mixing operation, so the L2 distance between the reconstructed frame $\mathsf{A}_n^t \otimes \mathsf{X}^t$ and the original frame $F^t$ should be minimized. 

%
%
%
%

\noindent\textbf{2) Contracting blocks: }
The contracting blocks of ICA-UNet are designed to further propagate the mixing tensor $\mathsf{A}_n^t$, 
and generate the learned ones in a multi-resolution manner. 
As shown in Fig.~\ref{fig:ica_unet},
the contracting part is made of $n$ contracting blocks, ranging from
$C_n$ to $C_{1}$. 
The contracting block $C_{k}$ ($k\!\in\![1,n]$) takes $\mathsf{A}_k^t$ as input, propagates it through a downsampling module (i.e., convolution with stride 2) and convolution modules (i.e., conv) sequentially, and outputs $\mathsf{A}_{k-1}^t$,
which is then forwarded to 
the next contracting block $C_{k-1}$ as well as the corresponding expanding block $E_{k}$.
%
%
%
%
%
%

\noindent\textbf{3) Expanding blocks: }
The expanding blocks $E_{k}$ ($k\!\in\![1, n\!-\!1]$) are designed to process the features generated by the contracting blocks and forward the outputs to ICA-decoder blocks and 
the next expanding block $E_{k+1}$ (or concatenation block $T_n$ for $E_{n-1}$). 
An expanding block has two sub-tasks:
upsampling the mixing tensor $\mathsf{A'}_{k}^t$ to $\mathsf{A'}_{k+1}^t$ (by upsample block $U_{k}$), and calculating dimension aligned correlation features $\mathsf{A'}_{c,k}^t$ from the neighbour frames (by concatenation block $T_{k}$).
Note we use $\mathsf{A'}_k^{t}$ to represent the mixing tensor in the upsampling path, distinguishing it from those in the downsampling path $\mathsf{A}_k^{t}$.

We use transposed convolution to achieve the upsampling on $\mathsf{A'}_k^{t}$, during which we obtain the outputs with various resolutions.
Taking $\mathsf{A}_n^{t-1}$, $\dots$, $\mathsf{A}_1^{t-1}$, and $\mathsf{A}_n^{t+1}$, $\dots$,  $\mathsf{A}_1^{t+1}$ from the outputs of contracting blocks for frames $F^{t-1}$ and $F^{t+1}$, respectively, we can 
calculate 
the temporal correlation features following \cite{ilg2017flownet} between mixing tensors $\mathsf{A}_k^{t-1}$ and $\mathsf{A}_k^{t}$, and between $\mathsf{A}_k^{t+1}$ and $\mathsf{A}_k^{t}$ ($k\!\in\![1, n]$).
The obtained correlation features explicitly
provide matching information from the neighbouring frames 
for more accurate segmentation.
The correlation features are then concatenated with $\mathsf{A'}_{k}^t$, 
and forwarded through a convolution module with $1\!\times\!1$ kernel (conv-bn-leakyrelu) for dimension reduction.
These computations are processed in the concatenation block $T_{k}$ as shown in Fig.~\ref{fig:ica_unet}.

\noindent\textbf{4) ICA-decoder: }
The ICA-decoder block is designed to mimic the mixing operation between the concatenated mixing tensor $\mathsf{A'}_{c,k}^t$ ($k\!\in\![1,n]$) and the basis $\mathsf{X}^t$, as in the standard ICA, to generate the output for evaluation.
As discussed earlier, transposed convolution is used as the mixing operation. 
The transposed convolution acts as both upsampling for evaluation and the multiplication projection field between $\mathsf{X}^t$ and each value in $\mathsf{A'}_{c,k}^t$, which helps reducing both the 
parameter size and the computation load.

After mixing, the mixed features are 
propagated through a convolution module as the output for evaluation.
From $E_{1}$ to $E_{n-1}$ we can obtain a total of $n\!-\!1$ multi-resolution segmentation outputs for evaluations, denoted as ${y'}_k$, $k\!\in\![1, n\!-\!1]$.
After we get $\mathsf{A'}_n^t$ from $E_{n-1}$, we forward $\mathsf{A'}_n^t$, $\mathsf{A}_n^{t+1}$, $\mathsf{A}_n^{t}$, $\mathsf{A}_n^{t-1}$, and $\mathsf{X}^{t}$ to the final concatenation block $T_n$ and its corresponding ICA-decoder, 
where we obtain the output with the same size as the original input.
Thus, we obtain a total of $n$ outputs in multi-resolution for evaluation.

\noindent\textbf{Objective function: }
We evaluate the outputs from the decoder blocks with the multi-resolutions ground-truth.
The overall objective function $\mathcal{L}$ is
\begin{equation}
    \mathcal{L} = \sum_{k=1}^{n}\alpha_k \mathcal{L}_{k}+\beta \mathcal{L}_{\text{ICA}}, k\in[1, n],
\end{equation}
where $\mathcal{L}_{k}$ ($k\!\in\![1, n]$) is the 
evaluation loss of the multi-resolution outputs in the decoder (with Cross-entropy);
$\alpha_k$ is the corresponding loss weights, 
while we take $\alpha_k=0.1$ for $k\!\in\![1, n\!-\!1]$
and $\alpha_n=1$;
$\mathcal{L}_{\text{ICA}}$ is the loss from Equation~\eqref{eq:icanet_loss}, 
and the weight $\beta$ is set to $0.2$ in our experiments.
Cross-entropy is used for calculating $\mathcal{L}_{k}$ with the rescaled versions of
ground truth.

\subsubsection{Latency reduction analysis}
For the inference of a frame of size $(1,1,d,h,w)$, the mixing tensor $\mathsf{A}_n^t$, 
as the input to the backbone, is composed of 
the $m$ coefficient tensors, each of size $(1,1,d/2,h/4,w/4)$, which is $32\times$ smaller than the original frame size $(1,1,d,h,w)$ for a regular U-Net. 
In addition, these coefficient tensors can be handled in parallel (i.e., a new task parallelism) due to the independence between the bases. 
Note that the processing of each coefficient tensor can still utilize any existing parallelization techniques such as model or operator parallelization \cite{gholami2018integrated,dryden2019improving,chetlur2014cudnn,vasudevan2017parallel} by applying them on the backbone. Therefore, significant latency reduction can be achieved.


\section{Experiments}
\noindent\textbf{Experiment Setup: }
We evaluate our model on an extended ACDC MICCAI 2017 challenge dataset made available
by MSU-Net \cite{wang2019msu} with labels on all the frames in the training data.
We compare our ICA-UNet with MSU-Net, the state-of-the-art real-time cine MRI segmentation method on the same dataset \cite{wang2019msu}.
We perform 5-fold cross-validation and use the average Dice score (the higher the better) to evaluate the segmentation accuracy.  
To further see how the accuracy of ICA-UNet compare against the state-of-the-art offline segmentation  
methods that achieves high accuracy but looses real-time performance, we evaluate the test data by submitting the segmentation results of ED and ES instants to ACDC online evaluation platform \cite{acdcchallenge}. 

All the methods were implemented in PyTorch and trained from scratch with the same hyperparameters and optimizer setting. MSU-Nets were based on the implementations by \cite{wang2019msu}. All the networks are fully parallelized using CUDA/CuDNN\cite{chetlur2014cudnn}. All experiments run on a machine with 16 cores of Intel Xeon E5-2620 v4 CPU, 256G memory, and an NVIDIA Tesla P100 GPU. 


\begin{table}[!tb]
\centering
\caption{Comparison of Dice, throughput (TP) and latency (LT) between ICA-UNets and the state-of-the-art real-time 3D cardiac cine MRI segmentation method MSU-Net. For ICA-UNet, $n$ denotes the number of  
contracting blocks. To satisfy real-time requirement, throughput should be above
22 FPS and the latency should be below 50 ms to avoid visually noticeable lags.}
\begin{tabular}{lcccccccc}
\toprule
\multirow{2}{*}{\textbf{Methods}} &  &  \multicolumn{4}{c}{\textbf{Dice score}} &\multirow{2}{*}{} & \textbf{TP} & \textbf{LT}\\
\cline{3-6}   
& & \textbf{RV} & \textbf{MYO} & \textbf{LV}  & \textbf{Average} & & (FPS) & (ms)\\ \midrule
MSU-Net(span=10) &  & .837$\pm$.034 & .811$\pm$.049 & .854$\pm$.040 & .834$\pm$.020 & & 70.2 & 442\\ 
MSU-Net(span=5)  &  & .855$\pm$.026 & .836$\pm$.022 & .897$\pm$.017 & .862$\pm$.011 & & 43.2 & 249\\
MSU-Net(span=3)  &  & .858$\pm$.034 & .838$\pm$.039 & .898$\pm$.034 & .864$\pm$.030 & & 29.4 & 169\\
MSU-Net(span=2) & & .860$\pm$.017 & .837$\pm$.031 & .901$\pm$.021 & .867$\pm$.020 & & 21.7 & 125\\
\midrule
ICA-UNet(n=3) &  & .900$\pm$.023 & .869$\pm$.027 & .934$\pm$.013 & .901$\pm$.017 & & 31.6 & \textbf{35}\\ 
ICA-UNet(n=4) &  & \textbf{.921}$\pm$\textbf{.017} & \textbf{.888}$\pm$\textbf{.034} & \textbf{.952}$\pm$\textbf{.015} & \textbf{.920}$\pm$\textbf{.019} & & 28.3 & 39\\ 
\bottomrule
\end{tabular}
\label{table:cine_result}
\end{table}
\begin{table}[!tb]
\centering
\caption{The Dice scores and Hausdorff distances on ACDC test set by ICA-UNet and state-of-the-art offline segmentation methods GridNet \cite{zotti2018convolutional}, ensemble U-Net \cite{isensee2017automatic},
{$\Omega$}-Net \cite{vigneault2018omega}. All results are reported by ACDC evaluation platform \cite{acdcchallenge}. ICA-UNet achieves comparable accuracy while satisfying real-time requirements.}
\begin{tabular}{lccccccccc}
\toprule
\multirow{2}{*}{\textbf{Methods}} &  &  \multicolumn{3}{c}{\textbf{Dice score}} &\multirow{2}{*}{} & \multicolumn{3}{c}{\textbf{Hausdorff (mm)}}\\
\cline{3-5} \cline{7-9} 
& & \textbf{RV} & \textbf{MYO} & \textbf{LV}  & & \textbf{RV} & \textbf{MYO} & \textbf{LV}\\ \midrule
Ensemble U-Net  &  & 0.923 & 0.911 & 0.950 & & 11.13 & 8.69 & 7.15 \\
$\Omega$-Net &  & 0.920 & 0.891  & 0.954  &  & N/A & N/A & N/A \\
GridNet &  & 0.910 & 0.894  & 0.938  &  & 11.80 & 9.45 & 7.30 \\
\midrule
ICA-UNet(n=4) &  & 0.920 & 0.890 & 0.940 & & 11.91 & 7.93 & 6.85 \\ 
\bottomrule
\end{tabular}
\label{table:test}
\end{table}



\noindent\textbf{Performance on real-time segmentation: }
The results of ACDC 3D cardiac cine MRI segmentation are shown in
Table~\ref{table:cine_result}. 
We can see that ICA-UNet increases the Dice score by 0.061(RV), 0.050(MYO), 0.051(LV), and 0.053(average), respectively, compared with the best results 
achieved by  MSU-Nets. ICA-UNets also achieve smaller Dice score variations than MSU-Nets in most cases.
In terms of throughput, although both ICA-UNets and MSU-Nets can satisfy the real-time requirement of 22 FPS,
only ICA-UNets can meet real-time latency requirement (below 50 ms), 
up to 12.6$\times$ faster than MSU-Nets. 
In summary, 
ICA-UNet not only achieves the best Dice score, but also is the only real-time segmentation method that can simultaneously meet the real-time 
throughput and latency requirements for visual guidance of cardiac interventions.

\begin{figure}[ht]
    \begin{center}
    \includegraphics[width=0.9\linewidth]{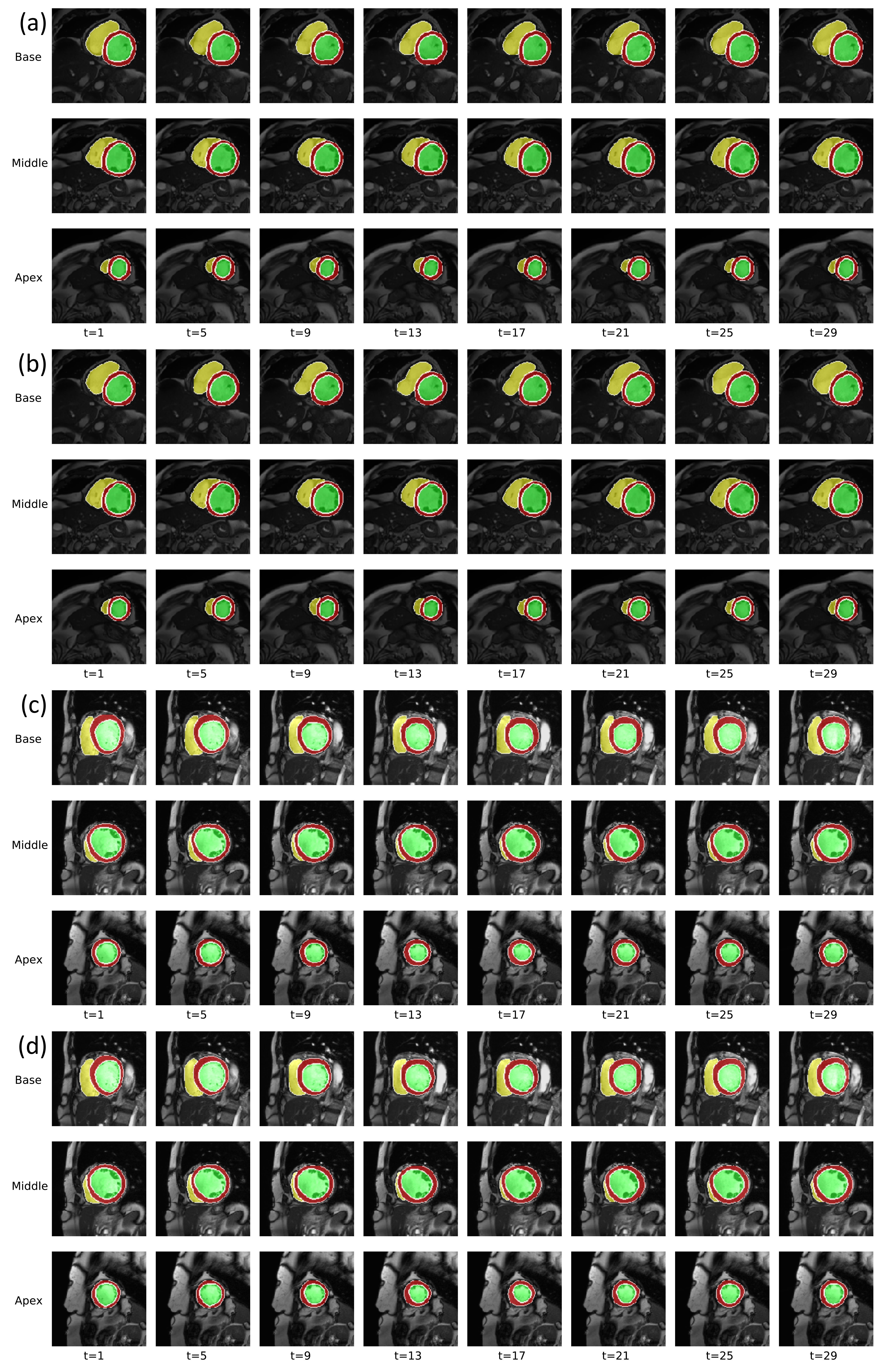}
    \end{center}
    \caption{The segmentation outputs of ICA-UNet (n=4) on 3D cardiac MRI cine for (a) patient 1 prediction, (b) patient 1 ground truth, (c) patient 2 prediction, and (d) patient 2 ground truth.  
    The rows indicate the slices at the bases, the middle, and the apex of LV. The columns show the results at various time steps in series. RV, MYO, and LV are labeled in yellow, red and green, respectively.}
    \label{fig:sup_patients}
\end{figure}

From the table, we can also see that the number of convolutional decoder blocks is an effective tuning 
knob for Dice and speed tradeoff. A higher number of blocks result in higher Dice scores at the cost of 
slightly reduced throughput and increased latency.
Visualization of segmentation results by ICA-UNet along with the corresponding ground truth is shown as Fig.~\ref{fig:sup_patients}.

\noindent\textbf{Accuracy v.s. state-of-the-art offline methods: }
To see how the accuracy of ICA-UNet compares with state-of-the-art offline segmentation methods which do not satisfy real-time requirements, we further verify our ICA-UNet on ED and ES instants of ACDC test data. The  evaluation results reported by \cite{acdcchallenge}, in terms of both dice score and Hausdorff distance, are shown in Table~\ref{table:test}.  
The results from the best approaches in the literature, including GridNet \cite{zotti2018convolutional}, {$\Omega$}-Net \cite{vigneault2018omega}, and ensemble U-Net \cite{isensee2017automatic}, are also included for reference. With complex network structures, the latency and throughput of these methods are far from the real-time  requirements, as shown in \cite{wang2019scnn}.  
In contrast, we see that the accuracy of ICA-UNet comes very close to these state-of-the-art results while meeting the real-time throughput and latency requirements.

\section{Conclusions}
Inspired by ICA, ICA-UNet decomposes temporal frames in 3D cardiac cine MRI into independent bases and the corresponding coefficient tensors, which are much smaller in size and help to learn better. Experimental results show that compared with the state-of-the-arts, ICA-UNet 
is the only 3D cardiac cine MRI segmentation method that can satisfy both real-time throughput and 
latency requirements with comparable (if not better) accuracy.


\bibliographystyle{splncs04}
\bibliography{bib}

\end{document}